\documentclass{amsart}
\usepackage{amsmath,amssymb,amsthm}
\usepackage{paralist}
\usepackage{graphics} 
\usepackage{epsfig}

\theoremstyle{plain}
\newtheorem{theorem}{Theorem}
\theoremstyle{remark}
\newtheorem{remark}{Remark}

\usepackage[colorlinks=true]{hyperref}
\hypersetup{urlcolor=blue, citecolor=red}

\newcommand{\R}{\mathbb{R}}

\begin{document}

\title{Simulating Nonholonomic Dynamics}

\author[M. Kobilarov]{Marin Kobilarov}
\address{M. Kobilarov: California Institute of Technology, Control and Dynamical Systems, Pasadena, CA 91125, USA}
\email{marin@cds.caltech.edu}

\author[D. Mart\'{\i}n de Diego]{David Mart\'{\i}n de Diego}
\address{D. Mart\'{\i}n de Diego: Instituto de Ciencias Matem\'aticas, CSIC-UAM-UC3M-UCM,
  Serrano 123, 28006,
  Madrid, Spain}
\email{david.martin@icmat.es}

\author[S. Ferraro]{Sebasti\'an Ferraro}
\address{S. Ferraro: Departamento de Matem\'atica e Instituto de Matem\'atica,
  Universidad Nacional del Sur, Av.\ Alem 1253,
  8000 Bah\'\i a Blanca, Argentina}
\email{sferraro@uns.edu.ar}


\begin{abstract}
  This paper develops different discretization schemes for
  nonholonomic mechanical systems through a discrete geometric
  approach. The proposed methods are designed to account for the
  special geometric structure of the nonholonomic motion. Two
  different families of nonholonomic integrators are developed and
  examined numerically: the geometric nonholonomic integrator (GNI) and the
  reduced d'Alembert-Pontryagin integrator (RDP). As a result, the
  paper provides a general tool for engineering applications, i.e. for
  automatic derivation of numerically accurate and stable dynamics
  integration schemes applicable to a variety of robotic
  vehicle models.
\end{abstract}

\maketitle

\section{Introduction}

Nonholonomic constraints have been the subject of deep analysis  since
the dawn of Analytical Mechanics.  Hertz, in 1894,  was the first to
use the term ``nonholonomic system'', but we can even find older
references  in the work by Euler in 1734, who studied the  dynamics of
a rolling rigid body moving without slipping on a horizontal plane.
Many authors have recently shown a new interest in that theory and
also in its relation  to the new developments in control theory,
subriemannian geometry, robotics, etc (see, for instance,
\cite{NF}). The main characteristic of this period  is that Geometry
was used in a systematic way (see L.D. Fadeev and A.M. Vershik
\cite{VF} as an advanced and fundamental reference, and also,
\cite{Bl2003,BlKrMaMu1996,CCMD,Cort,Koiller,LD} and references
therein).

In the case of nonholonomic mechanics, these constraint functions are,
roughly speaking, functions on the velocities that are not derivable
from position constraints.
Traditionally, the equations of motion for nonholonomic mechanics
are derived from the  Lagrange-d'Alembert principle which restricts
the set of infinitesimal variations (or constrained forces) in terms
of the constraint functions.

Recent works, such as \cite{Cortes_Martinez:Non-holonomic_integrators,
Fedorov_Zenkov:Discrete_nonholonomic_LL_systems_on_Lie_groups,IgMaDeMa2008,
McPe2006}, have introduced numerical integrators for nonholonomic
systems with very good energy behavior and properties such as the
preservation of the discrete nonholonomic momentum map. In this paper,
we will review and compare two new methods for nonholonomic mechanics,
the Geometric Nonholonomic Integrator (GNI)  \cite{FeIgDe2008} and
the Reduced d'Alembert-Pontryagin Integrator (RDP)  \cite{KoMaSu},
examining their behavior in the numerical simulation of some of the
most typical examples in nonholonomic mechanics: the Chaplygin sleigh
and the snakeboard.

Finally, the developed algorithms are packaged as a general computational
tool for automatic derivation of nonholonomic integrators given the system
constraints and Lagrangian. It is available for download from\\
\url{http://www.cds.caltech.edu/~marin/index.php?n=nhi}

\section{Introduction to  Discrete Mechanics}\label{1.2}
Discrete variational integrators appear as a special kind of geometric
integrators (see \cite{Hair,Sanz}). These integrators have  their
roots in the optimal control literature  in the 1960's  and 1970's. In
the sequel we will review the construction of this specific type of
geometric integrators (see \cite{Mars6} for an excellent survey about
this topic).

A discrete  Lagrangian
is a map $L_d\colon  Q \times Q\rightarrow \R$, where $Q$ is a
finite-dimensional configuration manifold. For the construction of
numerical integrators for a continuous Lagrangian system given by  a
Lagrangian $L: TQ\rightarrow \R$, the  discrete Lagrangian may be
considered as an approximation of the integral action
\[
L_d(q_0, q_1)\approxeq \int_0^h L(q(t), \dot{q}(t))\; dt
\]
where $q(t)$ is a solution of the Euler-Lagrange equations
corresponding to $L$, that is,
\begin{equation}\label{E-L}
  \frac{d}{dt}\left( \frac{\partial L}{\partial \dot{q}}(q(t),
  \dot{q}(t))\right)-\frac{\partial L}{\partial {q}}(q(t),
  \dot{q}(t))=0\; ,
\end{equation}
additionally satisfying $q(0)=q_0$ and $q(h)=q_1$, where $h$ is the
time step. Observe that this solution always exists if the Lagrangian
is regular and $h$ is small enough (see \cite{Patrick}).

Define the action sum $S_d: Q^{N+1}\rightarrow \R$ corresponding to
the Lagrangian $L_d$ by
\[
  S_d = \sum_{k=1}^{N}  L_d(q_{k-1}, q_{k}) ,
\]
where $q_k\in Q$ for $0\leq k\leq N$. For any covector $\alpha\in
T_{(x_1,x_2)}^*(Q\times Q)$, we have a decomposition
$\alpha=\alpha_1+\alpha_2$ where $\alpha_i\in T^*_{x_i} Q$. Therefore,
\[
dL_d(q_0, q_1)=D_{1} L_d(q_0, q_1)+D_{2} L_d(q_0,  q_1)\; .
\]
The discrete variational principle  states that the solutions of the
discrete system determined by $L_d$ must extremize the action sum
given fixed points $q_0$ and $q_N$.

Extremizing ${S_d}$ over $q_k$, $1\leq k\leq N-1$, we obtain the
following system of difference equations
\begin{equation}\label{Euler-Lagrange}
 D_1L_d(q_k, q_{k+1})+D_2L_d( q_{k-1}, q_{k})=0\; .
\end{equation}
These  equations are usually called the {\em discrete Euler-Lagrange
  equations}.

The geometrical properties corresponding to this numerical method are
obtained defining two discrete Legendre transformations associated to
$L_d$ by
\[
\begin{array}{rrcl}
  \mathbb{F}^- L_d:& Q\times Q&\longrightarrow& T^*Q \\
  & (q_0, q_1)&\longmapsto & (q_0, -D_1 L_d(q_0, q_1))
\end{array}
\]
\[
\begin{array}{rrcl}
  \mathbb{F}^+ L_d:& Q\times Q&\longrightarrow& T^*Q\\
  &(q_0, q_1)&\longmapsto & (q_0, D_2 L_d(q_0, q_1))
\end{array}
\]
and the 2-form $\omega_d=(\mathbb{F}^\pm L_d)^*\omega_Q$, where
$\omega_Q$ is the canonical symplectic form on $T^*Q$.
We will say that the discrete Lagrangian is regular if $\mathbb{F}^-
L_d$ is a local diffeomorphism. We will have that:
\begin{eqnarray*}
  \hbox{$\mathbb{F}^- L_d$ is a local diffeomorphism }&\Leftrightarrow& \hbox{$\mathbb{F}^+ L_d$ is a local diffeomorphism}\\
  &\Leftrightarrow& \hbox{$\omega_d$ is symplectic}
\end{eqnarray*}
Under this regularity condition, this implicit system of difference
equations (\ref{Euler-Lagrange}) defines a local discrete flow
$
\Upsilon:  U\subset Q\times Q\longrightarrow Q\times Q$, by
$\Upsilon(q_{k-1}, q_k)=(q_k, q_{k+1})$.
The discrete algorithm determined by $\Upsilon$ preserves the
symplectic form $\omega_d$, i.e., $\Upsilon^*\omega_d=\omega_d$.
Moreover, if the discrete Lagrangian is invariant under the diagonal
action of a Lie group $G$, then the discrete momentum map $J_d:
Q\times Q \rightarrow {\mathfrak g}^*$ defined by
$
\langle J_d(q_k, q_{k+1}), \xi\rangle=\langle D_2L_d(q_k, q_{k+1}), \xi_Q(q_{k+1})\rangle
$
is preserved by the discrete flow. Here, $\xi_Q$ denotes the
fundamental vector field determined by $\xi\in {\mathfrak g}$:
\[
\xi_Q(q)=\frac{d}{dt}\Big|_{t=0}(\hbox{exp}(t\xi)\cdot q)\; .
\]

Therefore, these integrators are {\bf symplectic-momentum preserving
  integrators}.

In \cite{MaMaMa} we have obtained  a geometric derivation of
variational integrators that is also valid  for reduced systems (on
Lie algebras, quotient of tangent bundles by a Lie group action, etc.)

\section{Description of the nonholonomic dynamics}
The presence of nonholonomic (or holonomic) constraints gives rise to forces.
Nonholonomic systems are described by the Lagrange-D'Alembert's
principle which prescribes the constraint forces induced by the given
nonholonomic constraints. In the following we will describe the
equations of motion of a nonholonomic system in terms of Riemannian
geometric tools (see \cite{CCMD}).

Let $Q$ be an $n$-dimensional differentiable manifold, with local coordinates
$(q^i)$, $1\leq i\leq n$.
Consider a mechanical Lagrangian system $L: TQ\to \R$ defined by $
L(v_q)=\frac{1}{2}{\mathcal G}(v_q, v_q) - V(q)$, $v_q\in T_qQ$ or, locally
\begin{equation}\label{conn}
  L(q,\dot{q})=\frac{1}{2}g_{ij}(q)\dot{q}^i\dot{q}^j-V(q)\; .
\end{equation}
Here ${\mathcal G}$ is a Riemannian metric on $Q$ (locally defined by
the symmetric,  positive definite matrix $(g_{ij}(q))_{1\leq i, j\leq
  n}$) and $V$ represents a potential function. We know that the
equations of motion for a Lagrangian system are (\ref{E-L}) which, in
the case of a mechanical Lagrangian system of the form  (\ref{conn}),
admits a nice expression in terms of standard Riemmanian geometric
tools:
\begin{equation*}
  \nabla _{\dot{c}(t)} \dot{c}(t) = - {\rm grad}~
  V(c(t))
\end{equation*}
where   $\nabla$ is the Levi--Civita connection associated to
${\mathcal G}$ and, in coordinates,  ${\rm grad}~
V(c(t))=g^{ij}\frac{\partial V}{\partial q^j}$ where $(g^{ij})$ is the
inverse matrix of $(g_{ij})$.

Assume that the system is subjected to nonholonomic constraints,
defined by a regular distribution  $\mathcal{D}$ on $Q$, with
$\hbox{rank }\mathcal{D} = n-m$. Locally the nonholonomic constraints
are described by the vanishing of $m$ independent functions
\[
\phi^a=\mu^a_i(q) \dot{q}^i, \quad 1\leq a\leq m\ \hbox{  (the
  ``constraint functions'')}.
\]
The Lagrange--d'Alembert principle states that the equations of motion
for a nonholonomic system determined by the two data $(L,
\mathcal{D})$ are:
\begin{align}
  \label{E-L-n}
  \frac{d}{dt}\left( \frac{\partial L}{\partial \dot{q}^i}(q(t), \dot{q}(t))\right)-\frac{\partial L}{\partial {q}^i}(q(t), \dot{q}(t))&=\lambda_a \mu^a_i(q(t))\; ,\\
  \mu^a_i(q(t)) \dot{q}^i(t)&=0\nonumber
\end{align}
where $\lambda_a$, $1\leq a\leq m$ are Lagrange multipliers to be determined.
Using the Levi-Civita connection we find an intrinsic equation for the
nonholonomic equations:
\begin{equation*}
  \nabla _{\dot{c}(t)} \dot{c}(t) = - {\rm grad}~
  V(c(t))+\bar{\lambda}(t),\quad \dot{c}(t) \in
  {\mathcal D}_{c(t)},
\end{equation*}
where    $\bar{\lambda}$ is a section of ${\mathcal D}^{\perp}$ along $c$.
Here ${\mathcal D}^{\perp}$ stands for the orthogonal complement
of ${\mathcal D}$ with respect to the metric ${\mathcal G}$.

In coordinates, defining the $n^3$ functions $\Gamma^k_{ij}$
(Christoffel symbols for $\nabla$) by
\[
\nabla_{\!\!\frac{\partial}{\partial
    q^i}}\,\frac{\partial}{\partial
  q^j}=\Gamma^k_{ij}\frac{\partial}{\partial q^k},
\]
we may rewrite the nonholonomic equations of motion as
\begin{align*}
  \ddot{q}^k(t)+\Gamma^k_{ij}(c(t))\dot{q}^i(t)\dot{q}^j(t)&=
  -g^{ki}(c(t))\frac{\partial V}{\partial q^i}+ \bar{\lambda}_a(t)g^{ki}(c(t))\mu^a_i(c(t))\; , 
  \\
  \mu^a_i(c(t))\dot{q}^i(t)&=0\; . 
\end{align*}

\section{Geometric Nonholonomic Integrator -- {\bf
    GNI}}\label{sec:geom_construction}

Given a nonholonomic system $(L, \mathcal{D})$ where $L$ is a
Lagrangian system of mechanical type (\ref{conn}), using the metric
${\mathcal G}$, we may consider the complementary projectors
\begin{align*}
  {\mathcal P}\colon &TQ\rightarrow{\mathcal D}\hookrightarrow TQ\\
  {\mathcal Q}\colon &TQ\rightarrow{\mathcal D}^{\perp}\hookrightarrow TQ
\end{align*}
and their duals considered as mappings from $T^*Q$ to $T^*Q$.

The Geometric Nonholonomic integrator (GNI, in the sequel)  for a
nonholonomic system only needs to fix a discrete Lagrangian $L_d:
Q\times Q\rightarrow \R$ to derive a numerical scheme, that is, it is
not necessary to discretize the nonholonomic constraints for this type
of integrator. The \textbf{\emph{discrete nonholonomic equations}} proposed in
\cite{FeIgDe2008} are
\begin{subequations}\label{eq:propuesta original}
  \begin{align}\label{eq:propuesta original:1}
    {\mathcal P}^*_{|q_k}( D_1 L_d(q_k, q_{k+1}))+
    {\mathcal P}^*_{|q_k} (D_2 L_d (q_{k-1},
    q_k))&=0 \\
    \label{eq:propuesta original:2}
          {\mathcal Q}^*_{|q_k}( D_1 L_d(q_k, q_{k+1}))-
          {\mathcal Q}^*_{|q_k} (D_2 L_d (q_{k-1},
          q_k))&=0.
  \end{align}
\end{subequations}
The first equation is the projection of the discrete
Euler--Lagrange equations to the dual of the constraint distribution
$\mathcal{D}$, while the second one can be interpreted as an
elastic impact of the system against $\mathcal{D}$.
This defines a unique discrete evolution operator if and only if the
Lagrangian $L_d$ is regular, in the sense of Section \ref{1.2}.

Define the pre- and post-momenta using the discrete
Legendre transformations:
\begin{align*}
  p^+_{k-1, k}&= \mathbb{F}^+ L_d(q_{k-1}, q_k)=(q_k,D_2L_d(q_{k-1}, q_k))\in T^*_{q_k}Q\\
  p^-_{k, k+1}&= \mathbb{F}^- L_d(q_k, q_{k+1})=(q_k,-D_1L_d(q_k, q_{k+1}))\in T^*_{q_k}Q.
\end{align*}

In these terms, equation~\eqref{eq:propuesta original:2} can be rewritten as
\[
\mathcal{Q}^*_{|q_k}\left(\frac{p^-_{k, k+1}+ p^+_{k-1, k}}{2}\right)=0
\]
which means that the average of post- and pre-momenta satisfies the
nonholonomic constraints.

We can also rewrite the discrete nonholonomic equations as a jump of
momenta:
\begin{equation}\label{eq:jump}
  p^-_{k, k+1}=\left({\mathcal P}^*- {\mathcal
    Q}^*\right)\big|_{q_k}(p^+_{k-1,k}).
\end{equation}

\paragraph{Reversibility.} Note that the map $\mathcal S=\mathcal P^*-\mathcal
Q^*$ is an involution,  that is $\mathcal S^{-1} = \mathcal S$. Therefore,
it acts equivalently in both directions, i.e.\ it creates a reversible and
symmetric flow. Furthermore, it can be expressed as
\[
\mathcal S(q) = U(q)D U^{-1}(q),
\]
where $D$ is a diagonal matrix with elements $\pm1$ corresponding to
the eigenvalues of $\mathcal S$ while $U$ is an invertible matrix with
columns the eigenvectors of $S$. Thus, the update~\eqref{eq:jump} can
be written as
\begin{align}\label{eq:refl}
  U^{-1}(q_k)p^-_{k, k+1} = D U^{-1}(q_k)p^+_{k-1,k}
\end{align}
based on which one can regard the momentum as either remaining unchanged
(corresponding to $+1$ eigenvalues) or being reflected (corresponding
to $-1$ eigenvalues) with respect to the basis defined by the mapping
$U^{-1}$.

\paragraph{Preservation Properties.}Suppose that $Q$ is a manifold on
which a Lie group $G$ acts. Define for each $q\in Q$
\begin{align}\label{eq:gq}
  \mathfrak{g}^q=\left\{ \xi\in \mathfrak{g} \,|\, \xi_Q(q)\in
  \mathcal{D}_q \right\},
\end{align}
where $\xi_Q(q)$ is the infinitesimal generator vector field
corresponding to $ \xi\in \mathfrak{g}$ at the point $q$. The bundle
over $Q$ whose fiber at $q$ is $\mathfrak{g}^q$ is denoted by
$\mathfrak{g}^\mathcal{D}$.  Define the discrete nonholonomic momentum
map $J^{\mathrm{nh}}_d\colon Q\times Q\to (\mathfrak{g}^\mathcal{D})^*$ as
in~\cite{Cortes_Martinez:Non-holonomic_integrators} by
\begin{align*}
  J^{\mathrm{nh}}_d(q_{k-1},q_k)\colon \mathfrak{g}^{q_k}&\to \mathbb{R}\\
  \xi&\mapsto \left\langle D_2L_d(q_{k-1},q_k),\xi_Q(q_k) \right\rangle.
\end{align*}
For any smooth section $\widetilde \xi$ of $\mathfrak{g}^\mathcal{D}$
we have a function $(J^{\mathrm{nh}}_d)_{\widetilde \xi}\colon Q\times
Q\to \mathbb{R}$, defined as $(J^{\mathrm{nh}}_d)_{\widetilde
  \xi}(q_{k-1},q_k)=J^{\mathrm{nh}}_d(q_{k-1},q_k)\left( \widetilde
\xi(q_k) \right)$.

If $L_d$ is $G$-invariant and $\xi\in \mathfrak{g}$ is a horizontal
symmetry (that is, $\xi_Q(q)\in \mathcal{D}_q$ for all $q\in Q$), then
the GNI preserves $(J^{\mathrm{nh}}_d)_\xi$ (see~\cite{FeIgDe2008} for
a proof).

In some cases of interest,  it is possible to obtain an integrator
preserving energy applying the following theorem (see
\cite{FeIgDe2008}):
\begin{theorem}\label{thm:preservation of energy}
  Let the configuration manifold be a Lie group with a bi-invariant
  Lagrangian and with an arbitrary distribution $\mathcal{D}$, and take
  a discrete Lagrangian that is left-invariant. Then the
  GNI~\eqref{eq:propuesta original} is energy-preserving.
\end{theorem}

\subsection{Nonholonomic version of the RATTLE and SHAKE methods}

Consider a continuous nonholonomic system determined by the mechanical
Lagrangian $L\colon \mathbb{R}^{2n}\to \mathbb{R}$:
\[
L({q}, \dot{q})=\frac{1}{2} \dot{q}^T M \dot{q} -V(q)
\]
(with $M$ a  constant, invertible matrix) and the constraints
determined by $\mu(q)\dot{q}=0$ where $\mu(q)$ is a $m\times n$ matrix
with $\hbox{rank } \mu=m$.

Consider now the symmetric discretization
\begin{align*}
  L_d(q_k, q_{k+1}) &= \frac{1}{2}h L\left(q_k, \frac{q_{k+1}-q_k}{h}\right)+\frac{1}{2}h L\left(q_{k+1}, \frac{q_{k+1}-q_k}{h}\right)\\
  &= \frac{1}{2h}\left(q_{k+1}-q_k\right)^T M\left(q_{k+1}-q_k\right)-\frac{h}{2}\left(V(q_k)+V(q_{k+1})\right)\,.
\end{align*}
After some straightforward computations we obtain that equations
(\ref{eq:propuesta original:1}) and (\ref{eq:propuesta original:1})
for the proposed nonholonomic discrete system are
\begin{subequations}\label{poiu}
  \begin{align}
    q_{k+1}-2q_k+q_{k-1}&=-h^2 M^{-1} \left(V_q(q_{k})+\mu^T(q_k)\lambda_k\right)\label{zxc1}\\
    0&=\mu(q_k)\left(\frac{q_{k+1}-q_{k-1}}{2h}\right)\label{zxc2},
  \end{align}
\end{subequations}
where $\lambda_k$ are Lagrange multipliers. We recognize this set of
equations as an obvious extension of the SHAKE method proposed by
\cite{Ryckaert} to the case of nonholonomic constraints.

The momentum is approximated by
${p}_k= M(q_{k+1}-q_{k-1})/2h$. Denoting $p_{k+1/2}=M(q_{k+1}-q_k)/h$,
equations (\ref{zxc1}) and (\ref{zxc2}) are now rewritten in the form
\begin{align*}
  p_{k+1/2}&={p}_k -\frac{h}{2}\left(  V_q(q_{k})+\mu^T(q_k){\lambda}_k\right),\\
  q_{k+1}&= q_k+hM^{-1}p_{k+1/2},\\  
  0&=\mu(q_k)M^{-1} {p}_k. 
\end{align*}

The definition of  ${p}_{k+1}$  requires the knowledge of $q_{k+2}$
and, therefore, it  is is natural to apply another step of the
algorithm (\ref{zxc1}) and (\ref{zxc2}) to avoid this
difficulty. Then, we obtain the new equations:
\begin{align*}
  {p}_{k+1}&={p}_{k+1/2} -\frac{h}{2}\left( V_q(q_{k+1})+\mu^T(q_{k+1}){\lambda}_{k+1}\right),\\
  0&=\mu(q_{k+1})M^{-1} {p}_{k+1}.
\end{align*}

The interesting result is that we obtain a natural extension of the
RATTLE algorithm for holonomic systems to the case of nonholonomic
systems. Unifying the equations above we obtain the following
numerical scheme
\begin{subequations}\label{wer}
  \begin{align}
    p_{k+1/2}&={p}_k -\textstyle\frac{h}{2}\left( V_q(q_{k})+\mu^T(q_k)\lambda_k\right),\label{azxc1111}\\
    q_{k+1}&= q_k+hM^{-1}p_{k+1/2},  \label{azxc1211}\\
    0&=\mu(q_k)M^{-1} {p}_k\label{azxc2211},\\
    {p}_{k+1}&={p}_{k+1/2} -\textstyle\frac{h}{2}\left(V_q(q_{k+1})+\mu^T(q_{k+1})\lambda_{k+1}\right),\label{azxc11111}\\
    0&=\mu(q_{k+1})M^{-1} {p}_{k+1}\label{azxc22211}.
  \end{align}
\end{subequations}
These equations allow us to take a triple  $(q_k,  p_k, {\lambda}_k)$
satisfying the constraint equations (\ref{azxc2211}), compute
$p_{k+1/2}$ using \eqref{azxc1111} and then $q_{k+1}$ using
\eqref{azxc1211}. Then, equations \eqref{azxc11111} and
\eqref{azxc22211} are used to compute the remaining components of the
triple $(q_{k+1},  p_{k+1}, {\lambda}_{k+1})$. It is clear, applying
Theorem \ref{thm:preservation of energy} that, in the case $V=0$, the
numerical method is energy preserving.

\begin{remark}\label{rem:initial_conditions}
  {\rm {}From this Hamiltonian point of view, we have shown that the initial
    conditions for  this numerical scheme are constrained in a natural
    way ($(q_0, {p}_0)$ with  $\mu(q_0)M^{-1} {p}_0=0$), that is, the
    initial conditions are exactly the same as those for  the
    continuous system. Additionally, we select $\lambda_0=0$ (see
    \cite{FeIgDe2008})}.
\end{remark}

In \cite{FeIgDe_toappear}, the following theorem is proven.
\begin{theorem}
The nonholonomic RATTLE method is globally second-order convergent.
\end{theorem}

\subsection{Projected Version of the Nonholonomic RATTLE}\label{sec:mf}
The proposed nonholonomic RATTLE method can be expressed without the
use Lagrangian multipliers by projecting the equations of motion onto
the constraint distribution through the projection $\mathcal{P}$
defined in~\S\ref{sec:geom_construction}.

Assuming that the Lagrangian is regular and that matrix $\mu$ is full
rank (i.e. rank $m$)  \eqref{poiu} can be reformulated as
\begin{subequations}\label{poiu_Rn}
  \begin{align}
    \mathcal{P}(q_k)^T M\left(q_{k+1} - 2q_k + q_{k-1}\right) &= -h^2 \mathcal{P}(q_k)^T
    V_q(q_{k}) \label{poiu_Rn_a} \\
    \mathcal{Q}(q_k)^TM\left(\frac{q_{k+1}-q_{k-1}}{2h}\right) &=
    0 , \label{poiu_Rn_b}
  \end{align}
\end{subequations}
where the $n \times n$ matrices $\mathcal{Q}$ and $\mathcal{P}$ represent both orthogonal projectors and have
rank $m$ and $(n-m)$, respectively, and are defined by
\begin{subequations}
  \begin{align}
    \mathcal{Q}(q) &=
    M^{-1}\mu(q)^T\left(\mu(q)M^{-1}\mu(q)^T\right)^{-1}\mu(q), \label{eq:fQ}
    \\
    \mathcal{P}(q) &= \operatorname{Id} - \mathcal{Q}(q) \label{eq:fP},
  \end{align}
\end{subequations}
where $\operatorname{Id}$ is the identity matrix.

Eqs.~\eqref{poiu_Rn} correspond to \eqref{eq:propuesta original} for
the case $Q=\mathbb{R}^n$ and
furthermore can be put in the ``momentum jump'' form by adding
\eqref{poiu_Rn_a} and \eqref{poiu_Rn_b} to get
\begin{align}
  q_{k+1} &= q_k +  \left(\operatorname{Id} - 2 M^{-1}\mathcal{Q}(q_k)^T
  M\right)(q_k-q_{k-1}) - h^2 M^{-1} \mathcal{P}(q_k)^T V_q(q_k).
\end{align}
For a more realistic example, we can add control inputs $u \in U\subset
\mathbb{R}^c$ acting in the basis defined by the $(n \times c)$ matrix $B(q)$
to obtain the following discrete equations:
\begin{align*}
  q_{k+1} &= q_k +  \left(\operatorname{Id} - 2 M^{-1}
  \mathcal{Q}(q_k)^T M\right)(q_k-q_{k-1}) + h^2 M^{-1} \mathcal{P}(q_k)^T f(q_k,u_k),
\end{align*}
where the forces $f:Q\times U \rightarrow T^*Q$ are given by $f(q,u) =
B(q)u - V_q(q)$.

In terms of momentum variables the integrator can be equivalently
expressed as
\begin{subequations}\label{phq}
  \begin{align}
    p_{k+1/2} &= \left(\operatorname{Id} - 2 \mathcal{Q}(q_k)^T \right)
    p_{k-1/2}  + h\mathcal{P}(q_k)^T f(q_k,u_k)\\
    q_{k+1} &= q_k + h M^{-1} p_{k+1/2}
  \end{align}
\end{subequations}
providing an update scheme $(q_k, p_{k-1/2})\Rightarrow (q_{k+1},
p_{k+1/2})$.

A remaining critical step in completing the algorithm is to establish the
link between the discrete variables $(q_k,p_{k+1/2})$ for $k=0,...,N$ used
in~\eqref{phq} and the continuous curve $(q(t),p(t))$. In that
respect one can regard $p_k=(p_{k-1/2} + p_{k+1/2})/2$ as an
approximation to the continuous momentum at time $t=kh$,
i.e. $p_k\approx p(kh)$. The pair $(q_k,p_k)$ satisfies the
nonholonomic constraint by definition and is related, following
from~\eqref{phq}, to the ``midpoint'' momenta through
\begin{subequations}\label{ppm}
  \begin{align}
    p_k &= \mathcal{P}(q_k)^T p_{k+1/2} - \frac{h}{2}\mathcal{P}(q_k)^T
    f(q_k, u_k), \label{ppma} \\
    p_k &= \mathcal{P}(q_k)^T p_{k-1/2} + \frac{h}{2}\mathcal{P}(q_k)^T
    f(q_k, u_k). \label{ppmb}
  \end{align}
\end{subequations}
These expressions can be used to determine proper variables
$(q_1,p_{1/2})$ to initialize the update~\eqref{phq} given continuous
initial conditions $(q_0,p_0)\approx (q(0), p(0))$. Since there is a
set of solutions $p_{1/2}$ satisfying~\eqref{ppm} for a given $p_0$
the most natural choice is to pick $p_{1/2}$ satisfying the
constraints at $q_0$. Therefore, the condition becomes
\begin{align*}
  p_{1/2} &= p_{0}  + \frac{h}{2}\mathcal{P}(q_0)^T f(q_0,u_0).
\end{align*}

In summary, given initial conditions $(q_0,p_0)$ satisfying the
constraints, the dynamics is evolved forwards to reach the final state
$(q_N,p_N)$, also in the constraint submanifold, after $N$ time steps
through
\begin{align}\label{gnirn}
  \begin{split}
    p_{1/2} &= p_{0}  + \frac{h}{2} \mathcal{P}(q_0)^T f(q_0,u_0), \\
    p_{k+1/2} &= \left(\operatorname{Id} - 2 \mathcal{Q}(q_k)^T \right)
    p_{k-1/2}  + h\mathcal{P}(q_k)^T f(q_k,u_k), \\
    q_{k+1} &= q_k + h M^{-1} p_{k+1/2}, \\
    p_{N} &= \mathcal{P}(q_N)^Tp_{N-1/2} + \frac{h}{2}\mathcal{P}(q_N) f(q_N,u_N),
  \end{split}
\end{align}
for  $k = 1,...,N-1$.

\section{Reduced d'Alembert-Pontryagin integrator-(RDP)}
In this section we consider a class of mechanical systems which, in
addition to nonholonomic constraints, also possess symmetries of
motion arising from conservation laws. The interplay between the
constraints and symmetries is linked to an intrinsic structure of the
state space associated with important properties of the dynamics. Our
goal in this section is to develop integrators that respect this
structure and lead to more faithful numerical representation.

In \S\ref{sec:geom_construction} we introduced the action of a
symmetry group $G$ and its relation the evolution of the system
momentum. Additional structure arises whenever the dynamics and
constraints are $G$-invariant  that permits the
construction of \emph{reduced} nonholonomic integrators
\cite{IgMaDeMa2008,KoMaSu}.

Following~\cite{Bl2003}, define the subspaces $\mathcal V_q$ and
$\mathcal S_q$
according to
\[
\mathcal V_q = \{\xi_Q(q)\ |\ \xi\in\mathfrak{g}\}, \qquad \mathcal S_q =\mathcal
D_q \cap \mathcal V_q.
\]
Practically speaking, the \emph{vertical} space $\mathcal V_q$
represents the space of tangent vectors parallel to symmetry
directions while $\mathcal S_q$ is the space of symmetry directions that
satisfy the constraints. Equivalently, $\mathcal{S}_q$ can be regarded
as the space generated by elements in $\mathfrak{g}^q$, as defined
in~\eqref{eq:gq}. The group $G$ is chosen so that the Lagrangian $L$
and distribution $\mathcal{D}$ are $G$-invariant. In addition, we make
the standard assumption (see~\cite{Bl2003,CeMaRa2001a}) that  $T_q Q=\mathcal
D_q + \mathcal V_q$, for each $q\in Q$.

Since our main interest is in a configuration space that is by
construction of the form $Q=M\times G$ we will restrict any further
derivations to the \emph{trivial bundle} case. Using coordinates
$(r,g)\in M\times G$ a basis for $\mathfrak{g}^q$
can be chosen as $\{e_b(r,g)\}$, for
$b=1,...,\operatorname{dim}(\mathcal{S})$. Since $\mathcal{D}$ is
$G$-invariant these elements can be expressed as
$e_b(r,g)=\operatorname{Ad}_ge_b(r)$, where $\{e_b(r)\}$ is the
body-fixed basis. We denote $\mathfrak{g}^r$ the space spanned by
$\{e_b(r)\}$ at each $(r,e)\in Q$. Lastly, the system is subject to
control force $f:[0,T]\rightarrow T^*M$ restricted to the shape
space.

\paragraph*{Nonholonomic Connection}
With these definitions we can define a principal connection
$\mathcal{A}:TQ\rightarrow\mathfrak{g}$ with horizontal distribution
that coincides with $\mathcal H_q$ at the point $q$, where
$\mathcal{D}_q=\mathcal S_q\oplus \mathcal H_q$. This connection is
called the \emph{nonholonomic connection} and is constructed according
to $\mathcal A =
\mathcal A^{\text{kin}} + \mathcal A^{\text{sym}}$, where $\mathcal
A^{\text{kin}}$ is the kinematic connection enforcing the nonholonomic
constraints and $\mathcal A^{\text{sym}}$ is the mechanical connection
corresponding to symmetries satisfying the constraints. These maps are
defined according to
\begin{align}\label{eq:Ar}
  \begin{split}
    \mathcal A^{\text{kin}}(q)\cdot \dot q &= 0, \\
    \mathcal A^{\text{sym}}(q)\cdot \dot q &=
    \operatorname{Ad}_g \Omega,
  \end{split}
\end{align}
where $\Omega\in\mathfrak g^r$ is called the \emph{locked angular
  velocity}, i.e. the velocity resulting from instantaneously locking
the joints described by the variables $r$. Intuitively, when the
joints stop moving the system continues its motion uniformly along a
curve (with tangent vectors in $\mathcal{S}$) with body-fixed velocity
$\Omega$ and a corresponding spatial momentum that is conserved.

By definition the principal connection can be expressed as
\[\mathcal A(q)\cdot \dot q =
\operatorname{Ad}_g(g^{-1}\dot g + \mathcal A(r)\dot r),\]
where $\mathcal{A}(r)$ is the local form and the two components
in~\eqref{eq:Ar} can be added to obtain
\[
g^{-1}\dot g + \mathcal A(r)\dot r = \Omega.
\]

\paragraph{Numerical Formulation}
Since the Lagrangian is $G$-invariant, we can define the \emph{reduced
  Lagrangian} $\ell:TM\times\mathfrak{g}\rightarrow\mathbb{R}$
\begin{align}\label{eq:ell}
  \ell(r, \dot r, \xi) = L(r, \dot r, e, g^{-1}\dot g).
\end{align}

In~\cite{KoMaSu} a nonholonomic integrator was derived using a
discrete variational d'Alembert-Pontryagin principle based on the
reduced Lagrangian $\ell$, the connection $\mathcal{A}$ and a chosen
trajectory discretization. In particular, a discrete trajectory with
points $q_k=(r_k,g_k)\in M\times G$ and respective velocities $u_k \in
TM$ and $\xi_k \in \mathfrak{g}$ was constructed so that
\[
r_{k+1}-r_k = hu_k, \qquad \tau^{-1}(g_k^{-1}g_{k+1})=h\xi_k,
\]
where $\xi_k=\Omega_k - \mathcal{A}(r_{k+\alpha})u_k$,
with $r_{k+\alpha} := (1-\alpha)r_k + \alpha r_{k+1}$ for a chosen
$\alpha\in[0,1]$. The map $\tau:\mathfrak{g}\rightarrow G$
represents the \emph{difference} between two configurations in the
group by an element in its algebra and can be selected as:
\begin{itemize}
\item Exponential map $\exp: \mathfrak{g} \rightarrow G$, defined by
  $\exp(\xi)=\gamma(1)$, with $\gamma: \mathbb{R} \rightarrow G$ is
  the integral curve through the identity of the left invariant vector
  field associated with $\xi \in \mathfrak{g}$ (hence, with
  $\dot{\gamma}(0)=\xi$);
\item Canonical coordinates of the second kind
  $\operatorname{ccsk}:\mathfrak{g} \rightarrow G$,
  $\operatorname{ccsk}(\xi)=\exp(\xi^1 e_1)\cdot\exp(\xi^2 e_2)
  \cdot...\cdot\exp(\xi^n e_n)$, where $\{e_i\}$ is the Lie algebra
  basis.
\end{itemize}%

A third choice for $\tau$, valid only for certain {\it quadratic}
matrix groups~\cite{CaOw2003} (which include the rigid motion groups
$\operatorname{SO}(3)$, $\operatorname{SE}(2)$, and
$\operatorname{SE}(3)$), is the \emph{Cayley map} $\operatorname{cay}
: \mathfrak{g} \rightarrow G$,
$\operatorname{cay}(\xi)=(e-\xi/2)^{-1}(e+\xi/2).$ (See App.~\ref{app:dtau} for more details).

With these definitions in place the resulting reduced
d'Alembert-Pontryagin (RDP) integrator can be stated~\cite{KoMaSu}.
For numerical convenience it is given in terms of vector-matrix notation,
by treating the Lie algebra variables $\xi$ and $\Omega$ as vectors of
coordinates with respect to a chosen canonical basis (see
App.~\ref{app:se2} for an example).

The discrete flow satisfies the \emph{reduced discrete dynamics}
\begin{align}\label{eq:rdp_dyn}
  \hspace{-10pt}  \left[\hspace{-5pt}
    \begin{array}{cc}
      \operatorname{Id} & \left[\mathcal{A}(r_k)\right]^T \\
      {0} & [e_1(r_k), ..., e_c(r_k)]^T
    \end{array}
    \hspace{-5pt}\right] \hspace{-3pt}\left( \hspace{-2pt}\left[\hspace{-5pt}
    \begin{array}{c}
      \partial_u \ell_k \\
      (\operatorname{d\tau}^{-1}_{h\xi_k})^*\partial_{\xi}\ell_k
    \end{array}
    \hspace{-5pt}\right] \hspace{-3pt} - \hspace{-3pt}\left[\hspace{-5pt}
    \begin{array}{c}
      \partial_u \ell_{k-1} \\
      (\operatorname{d\tau}^{-1}_{-h\xi_{k-1}})^*\partial_{\xi}\ell_{k-1}
    \end{array}
    \hspace{-5pt}\right]\hspace{-2pt} \right) \hspace{-2pt} = \hspace{-2pt} \left[\hspace{-5pt}
    \begin{array}{c}
      hf_k \\
      0
    \end{array}\hspace{-5pt}\right],
\end{align}
where $\ell_k := \ell(r_{k+\alpha},u_k,\xi_k)$ and $\xi_k = \Omega_k -
\mathcal A(r_{k+\alpha})u_k$.  The map
$\operatorname{d\tau}_{\xi}:\mathfrak{g}\rightarrow\mathfrak{g}$ is the {\it right-trivialized tangent} of $\tau(\xi)$ defined by
$\operatorname{D}\tau(\xi)\cdot\delta=TR_{\tau(\xi)}(\operatorname{d\tau}_{\xi}\cdot\delta)$
and $d\tau^{-1}_{\xi}:\mathfrak{g}\rightarrow\mathfrak{g}$ is its
inverse (see App.~\ref{app:dtau}).

Equation~\eqref{eq:rdp_dyn} along with the \emph{reconstruction equations}
\begin{align}\label{eq:rdp_rec}
    g_{k+1} = g_k\tau(h\xi_k), \qquad r_{k+1} = r_k + hu_k,
\end{align}
constitute the complete RDP discrete evolution.

\section{Examples}\label{sec:apps}

\subsection{The Chaplygin Sleigh}\label{sec:cs}
\begin{figure}[ht]
  \begin{center}\includegraphics[width=1.5in]{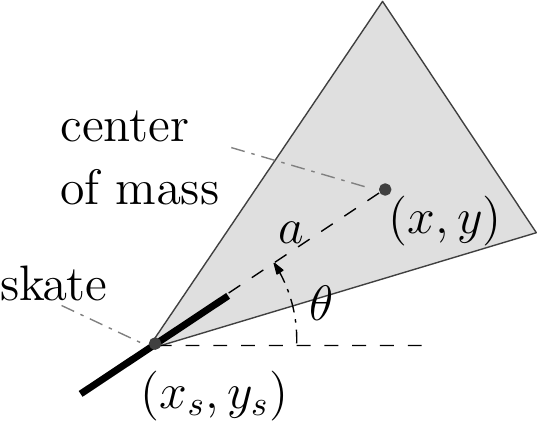}\end{center}
  \caption{Chaplygin Sleigh model.}
  \label{fig:cs}
\end{figure}

The Chaplygin Sleigh~\cite{Bl2003} is a planar rigid body making a
contact with the ground through  a \emph{skate} mounted at the central
axis of the body at a distance $a$ from its center of mass
(Fig.~\ref{fig:cs}). The configuration space is the group $G=\operatorname{SE}(2)$ with
coordinates $q=(\theta, x, y)$ describing the orientation and the
position of the center of mass. The body has rotational inertia $I$
and mass $m$ and, therefore, its Lagrangian is defined by
\begin{align}\label{eq:cs_L}
    L(q,\dot q) = \frac{1}{2}I\dot\theta^2 + \frac{1}{2}m(\dot x^2 +  \dot y^2).
\end{align}

At the point of the skate contact $(x_s, y_s) = (x - a \cos\theta, y - a
\sin \theta)$ the body must slide in the direction in which it is
pointing. This condition is encoded by the nonholonomic constraint
\[
a\dot \theta + \sin\theta \dot x - \cos\theta \dot y = 0.
\]
A structure-preserving integrator was developed in~\cite{Fedorov_Zenkov:Discrete_nonholonomic_LL_systems_on_Lie_groups}
based on the discrete Lagrange-d'Alembert (DLA) principle with
discrete momentum and measure preservation properties. Exploring
this direction further, in this section we develop two alternative
methods based on the GNI and RDP schemes.

\paragraph{GNI Integrator.} From the mass matrix $M=\text{diag}(I, m,
m)$ and the constraint $\mu_1(q) = [a, \sin\theta, -\cos\theta]$, the
projector $\mathcal Q$ can be computed using~\eqref{eq:fQ} as
\begin{align}
  \mathcal Q(q) = \frac{1}{I+a^2m}\left[\begin{array}{ccc}
      a^2m & am\sin\theta & -am\cos\theta \\
      aI\sin\theta & I \sin\theta^2 & -I\sin\theta\cos\theta \\
      -aI\cos\theta & -I\sin\theta\cos\theta & I\cos^2\theta
    \end{array}\right].
\end{align}

Since the mass matrix is constant, the GNI integrator can be derived
according to $v_{k+\frac{1}{2}} = (\text{Id} -
M^{-1}\mathcal Q(q_k)^T M)v_{k-\frac{1}{2}}$. In terms of the
coordinates $v=(v^\theta, v^x, v^y)$, the discrete update becomes
\begin{align*}
  \begin{split}
    & v_{k+\frac{1}{2}}^\theta = \left(1\!-\!\frac{2a^2m}{I'}\right)
    v_{k-\frac{1}{2}}^\theta + \frac{am}{I'} \left(-2\sin\theta_k v^x_{k-\frac{1}{2}}
    + 2\cos \theta_k v^y_{k-\frac{1}{2}} \right),
    \\
    & v_{k+\frac{1}{2}}^x = -\frac{2aI}{I'} \sin\theta_k v_{k-\frac{1}{2}}^\theta +
    \left(1\!-\!\frac{2I}{I'}\sin^2\!\theta_k\right)v_{k-\frac{1}{2}}^x + \frac{2I}{I'}\sin\theta_k\cos\theta_k v_{k-\frac{1}{2}}^y,
    \\
    & v_{k+\frac{1}{2}}^y = \frac{2aI}{I'}\cos\theta_k v_{k-\frac{1}{2}}^\theta + \frac{2I}{I'}\sin\theta_k
    \cos\theta_k v_{k-\frac{1}{2}}^x +
    \left(1\!-\!\frac{2I}{I'}\cos^2\theta_k \right)v_{k-\frac{1}{2}}^y,
  \end{split}
\end{align*}
where $I' = I +a^2m$. It is straightforward to verify that the
resulting update rule is energy-preserving, i.e. $\langle Mv_{k-1/2},
v_{k-1/2}\rangle = \langle M v_{k+1/2},v_{k+1/2}\rangle $. This
property is inherent to the GNI construction as explained
in~\cite{FeIgDe2008}.

\paragraph{RDP Integrator.} The sleigh has no internal joints and
therefore no shape space. Since the Lagrangian~\eqref{eq:cs_L} is
left-invariant to $\operatorname{SE}(2)$ group action, the reduced
Lagrangian~\eqref{eq:ell} can be expressed as
\[
\ell(\xi) = L(e, g^{-1}\dot g),
\]
where $\xi=(\omega, v, v^\perp) \in \mathfrak{g}$ describes the
angular, forward, and sideways velocities with respect
to the body frame fixed at the center of mass. The constrained
symmetry space~\eqref{eq:gq} of the sleigh can be identified as
\[
\mathfrak{g}^q = \text{span}\{e_1(g), e_2(g) \},
\]
where $e_1 = (1, 0, a) \in \mathfrak g$ and $e_2=(0, 1, 0) \in \mathfrak g$ form the constant basis in
the body-fixed frame with $e_i(g) = \text{Ad}_g e_i$, for
$i=1,2$. The two
components of the nonholonomic momentum $p_i = \langle\partial_\xi\ell
, e_1\rangle $ become
\begin{align*}
  p_1 = (J+a^2m) \omega, \qquad p_2 = m v,
\end{align*}
corresponding to angular and forward momenta, respectively. The group
trajectory can be reconstructed from the momentum according to
\[
g^{-1}\dot g = \left( \frac{1}{I + a^2m}p_1,\ \frac{1}{m}p_2,\
\frac{a}{I + a^2m}p_1 \right).
\]
The momentum components themselves evolve according to $\dot p_i =
\langle \text{ad}^*_\xi\partial_\xi\ell, e_i\rangle$ (see
~\cite{BlKrMaMu1996}), or equivalently
\[
 \dot p_1 = -\frac{a}{I + a^2m}p_1p_2, \qquad \dot p_2 = \frac{ma}{(I
   + a^2m)^2} p_1^2.
\]

Since the shape space consists of a single point, the discrete dynamics includes only the momentum
equations~\eqref{eq:rdp_dyn} which become
\[
\langle
(\operatorname{d\tau}^{-1}_{h\xi_{k}})^*\partial_{\xi}\ell_{k} -
(\operatorname{d\tau}^{-1}_{-h\xi_{k-1}})^*\partial_{\xi}\ell_{k-1},
e_i\rangle = 0,
\]
for $i=1,2$. A simple form of these equations can be derived by
choosing $\tau=\exp$ and truncating its tangent to first order,
i.e. $\operatorname{d\tau}^{-1}_{\xi} \approx \text{Id} -
\frac{1}{2}\text{ad}_\xi$. Using the notation $p_k=( (p_1)_k, (p_2)_k
)$ the update becomes
\begin{align*}\vspace{-5pt}
  &  (p_1)_k - (p_1)_{k-1} = -\frac{ha}{2(J+a^2m)}\left[(p_1)_k (p_2)_k + (p_1)_{k-1}(p_2)_{k-1}\right] ,\\
  &  (p_2)_k - (p_2)_{k-1} = \frac{hma}{2(J+a^2m)^2} \left[{(p_1)_k}^2 + {(p_1)_{k-1}}^2\right].\vspace{-5pt}
\end{align*}
These conditions are used to solve for the unknown next momentum $p_k$,
e.g. through cubic equation root-finding. Note, that this particular choice of
approximation exactly matches a standard implicit central difference
discretization of the continuous ODE. This is generally not
case for systems with non-constant Lie algebra basis element $e_i$
such as the snakeboard. Higher accuracy can be achieved through other
choices of $\tau$ and better approximation of $\operatorname{d\tau}$.
App.~\ref{app:se2} details the cases $\tau=\exp$ and
$\tau=\operatorname{cay}$ on $\operatorname{SE}(2)$.

The reconstruction equations are
\[
g_{k+1} = g_k\exp(h\xi_k),
\]
where
\[
\xi_k = \left( \frac{1}{I + a^2m}(p_1)_k,\ \frac{1}{m}(p_2)_k,\
\frac{a}{I + a^2m}(p_1)_k \right).
\]

\begin{figure}[h]
  \hspace{-15pt}\includegraphics[width=3in]{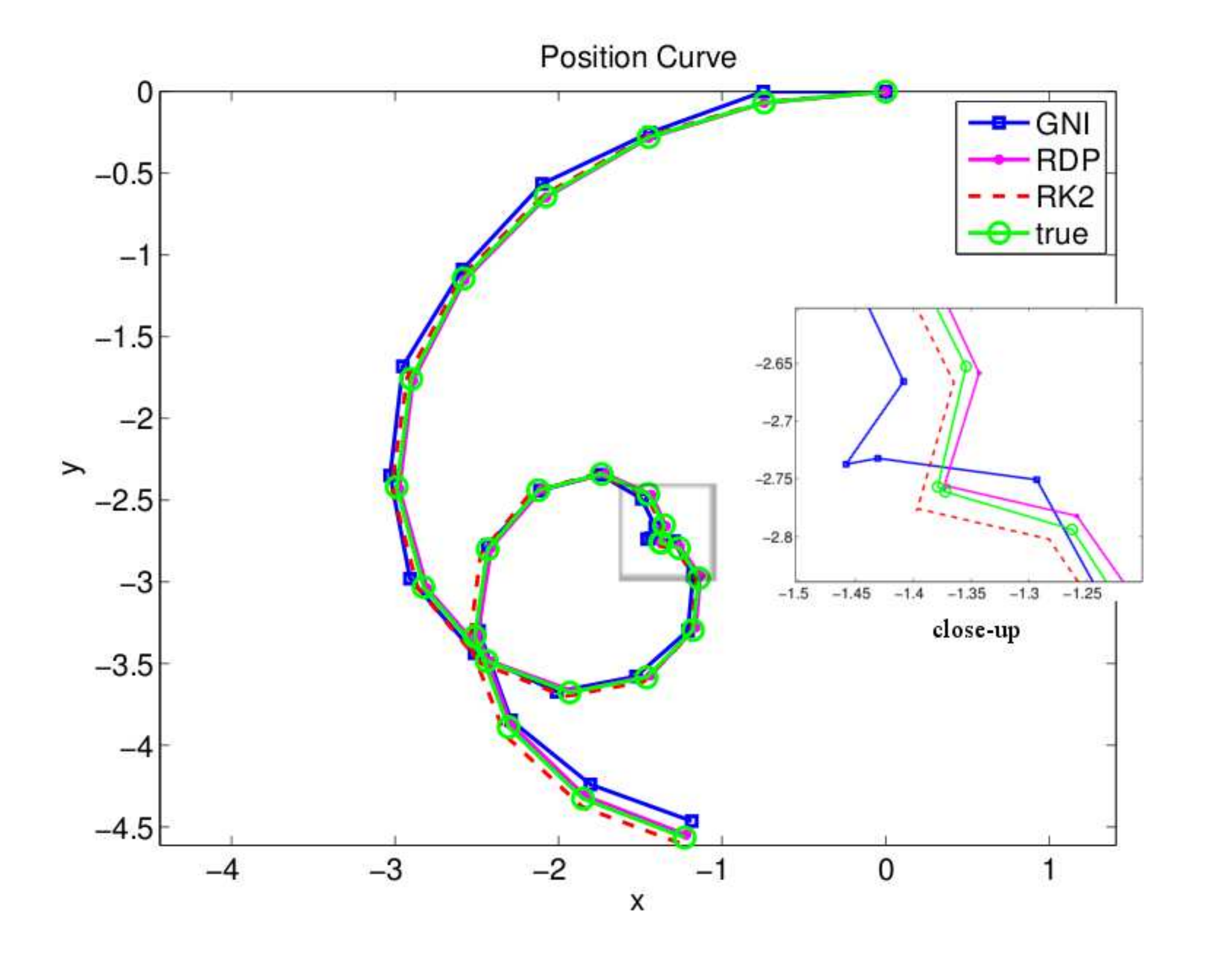}\vspace{-10pt}\includegraphics[width=2.2in]{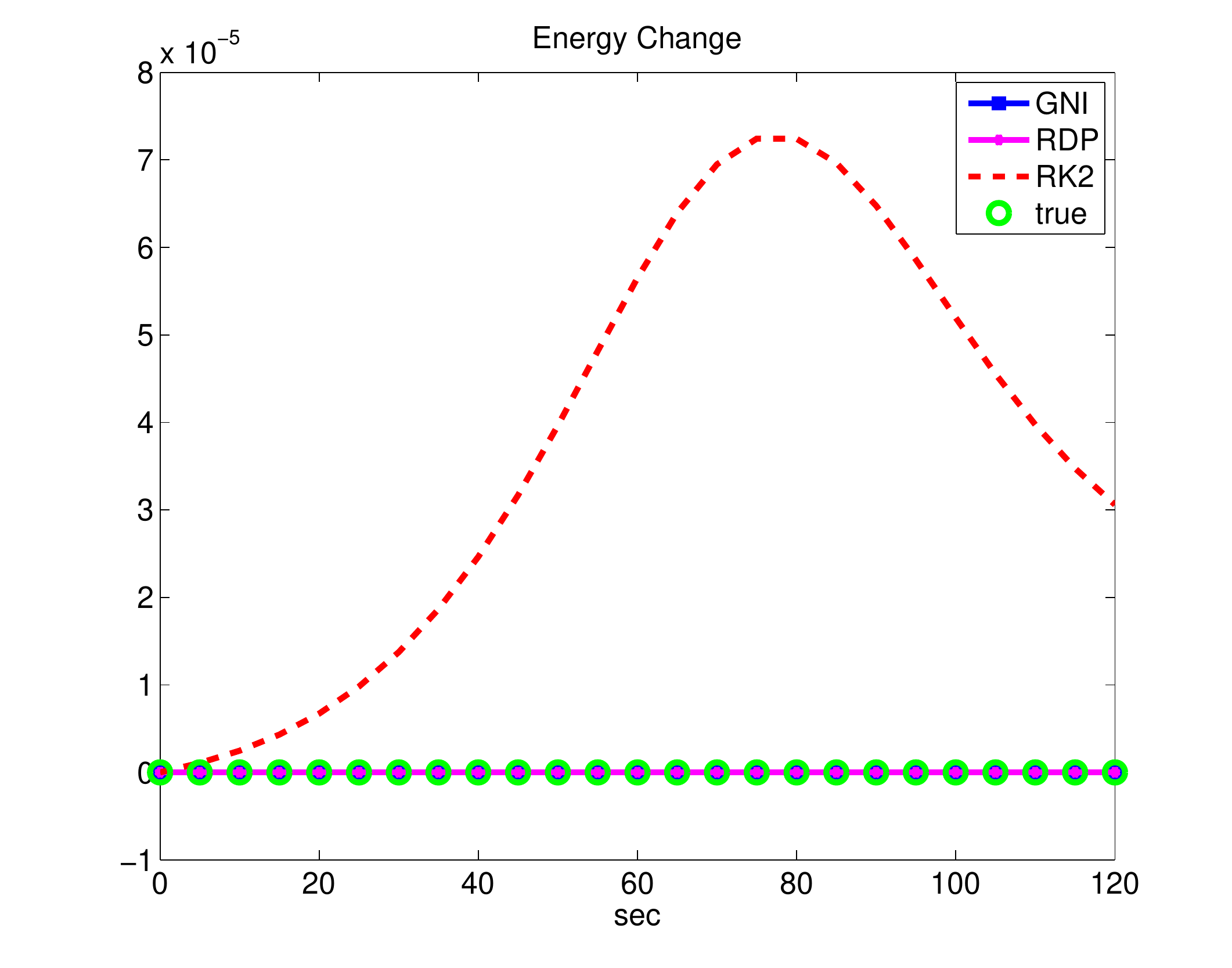}
  \caption{Position curves (left) of the sleigh integrators and the
    corresponding energy (right). The embedded close-up frame (left)
    zooms in on the cusp point of the ``heart'' shape. }
  \label{fig:cs_plots}
\end{figure}

\paragraph{Numerical Comparisons.} The numerical behavior of the
algorithms is now examined in terms of their ability to reproduce the
true system trajectory and in terms of their energy
preservation. Comparison to a standard Runge-Kutta second-order method
is also included.

Note that the standard Chaplygin sleigh model
(e.g. ~\cite{Bl2003,Fedorov_Zenkov:Discrete_nonholonomic_LL_systems_on_Lie_groups,FeIgDe2008}) is studied in terms of the
coordinates of the skate contact rather than the center
off mass as in this work. For easier reference to such previous
studies, we present the position curves below in terms of the skate
coordinates $(x_s,y_s)$. This representation enables the generation of
the familiar ``heart''-shaped curves (Fig.~\ref{fig:cs_plots}).

\subsection{The Snakeboard}\label{sec:sb}

\begin{figure}[h]
  \begin{center}\includegraphics[width=1.5in]{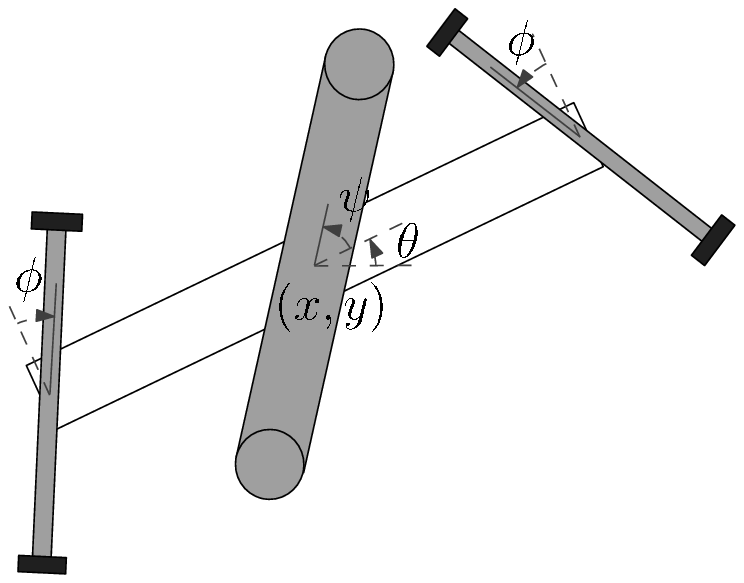}\hspace{-0pt}\includegraphics[width=3.2in]{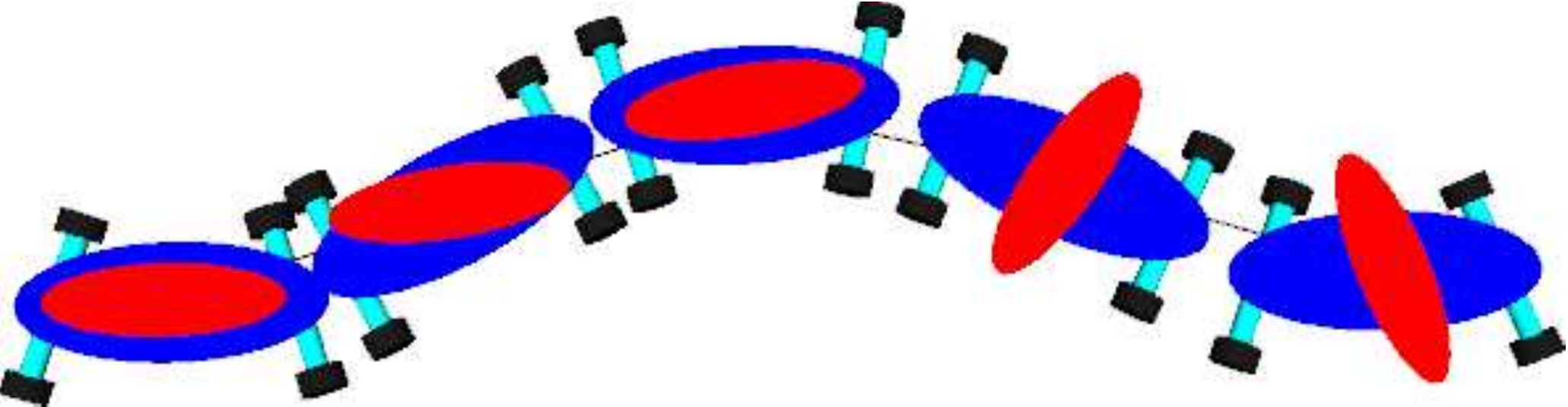}\end{center}
  \caption{Snakeboard model (left) and a typical trajectory (right).}
  \label{fig:sb}
\end{figure}

The snakeboard (Fig.~\ref{fig:sb}) represents a type of system with an
interesting interplay between constraints and symmetries. It has
served as a classical example (e.g.~\cite{BlKrMaMu1996,CeMaRa2001a,bullolewis})
of a system with non-trivial intersection of the constraint
distribution $\mathcal D$ and the vertical space $\mathcal V$. Our
integrators capture the dynamics of such systems and their performance
is examined in this section.

The shape space variables of the snakeboard are $r=(\psi, \phi)\in
\mathrm{S}^1\times \mathrm{S}^1$ denoting the rotor angle and the
steering wheels angle, while its configuration is defined by $(\theta,
x, y)$ denoting orientation and position of the board. This
corresponds to a configuration space $Q=\mathrm{S}^1\times \mathrm{S}^1 \times
\operatorname{SE}(2)$ with shape space $M=\mathrm{S}^1\times
\mathrm{S}^1 $ and group $G=\operatorname{SE}(2)$. Additional
parameters are its mass $m$, distance $l$ from its center to the
wheels, and moments of inertia $I$ and $J$ of the board and the
steering. The kinematic constraints of the snakeboard are:
\begin{align}\label{eq:sb_con}
  \begin{split}
    &-l\cos \phi
    d\theta -\sin(\theta+\phi) dx + \cos(\theta + \phi) dy= 0, \\
    &l\cos \phi d\theta -\sin(\theta-\phi) dx + \cos(\theta - \phi)dy =
    0,
  \end{split}
\end{align}
enforcing the fact that the system must move in the direction in which
the wheels are pointing and spinning. The constraint distribution is
spanned by three covectors:
\[
\mathcal{D}_q = \operatorname{span} \left\{ \frac{\partial}{\partial
    \psi}, \frac{\partial}{\partial \phi}, c\frac{\partial}{\partial
    \theta} + a\frac{\partial}{\partial x} + b\frac{\partial}{\partial
    y}\right\},
\]
where $a=-2 l\cos\theta \cos^2\phi, \; b=-2l\sin\theta \cos^2 \phi, \;
c=\sin 2\phi$. The group directions defining the vertical space are:
\[
\mathcal{V}_q = \operatorname{span} \left\{ \frac{\partial}{\partial
  \theta}, \frac{\partial}{\partial x}, \frac{\partial}{\partial
  y}\right\},
\]
and therefore the constrained symmetry space becomes:
\begin{align}
  \mathcal{S}_q = \mathcal{V}_q \cap \mathcal{D}_q =
  \operatorname{span} \left\{ c\frac{\partial}{\partial \theta} +
  a\frac{\partial}{\partial x} + b\frac{\partial}{\partial y}
  \right\}. \label{eq:Ss}
\end{align}
Since $\mathcal{D}_q = \mathcal{S}_q \oplus \mathcal{H}_q,$ we have
$\mathcal{H}_q = \operatorname{span} \left\{ \frac{\partial}{\partial
  \psi}, \frac{\partial}{\partial \phi}\right\}.$ Finally, the
Lagrangian of the system is $L(q, \dot q) = \frac{1}{2}\dot q^T M \dot
q $ where
\[
\operatorname{M}=\left[\begin{array}{ccccc}
    I & 0 & I & 0 & 0 \\
    0 & 2J & 0 & 0 &  0\\
    I & 0 & ml^2 & 0 & 0  \\
    0 & 0 & 0 & m & 0  \\
    0 & 0 & 0 & 0 & m
  \end{array}\right].
\]
The reduced Lagrangian can be expressed as $\ell(r, u, \xi ) = (u, \xi)^T
\operatorname{M}\;(u,\xi )$ by treating the velocity $\xi$ as a vector
in the standard $\mathfrak{se}(2)$ basis (defined in App.~\ref{app:se2}).

There is only one direction along which
snakeboard motions lead to momentum conservation: it is defined by the
basis element
\[ e_1(r) = 2l\cos^2\phi \left[ \begin{array}{c}\frac{\tan\phi}{l} \\
    -1 \\0 \end{array} \right], \] and, hence, there is only one
momentum variable $p_1 \!=\! \big\langle \frac{\partial
  \ell}{\partial \xi}, e_1(r)\big\rangle$. Using this variable we can
derive the connection according to~\cite{BlKrMaMu1996} as
\[ [\mathcal{A} ] = \left[ \begin{array}{cc}
    \frac{I}{ml^2}\sin^2 \phi & 0 \\
    -\frac{I}{2ml}\sin 2\phi & 0 \\
    0 & 0
  \end{array} \right]  , \text{and }\quad
\Omega = \frac{p_1}{4ml^2\cos^2\phi}e_1(r).
\]

\paragraph{GNI Integrator.}The snakeboard
constraints~\eqref{eq:sb_con} can be expressed in terms of the one-forms
\[
\mu_1(q) = (0, 0, a, -c, 0), \qquad \mu_2(q) = (0, 0, b, 0, -c).
\]
The projector $\mathcal Q$ can then be computed from $\mu$ and the
mass matrix $M$ using~\eqref{eq:fQ} to obtain
\begin{align*}
  \mathcal Q(q) = \frac{1}{ml^2\!-\!I\sin^2\phi}\left(\begin{array}{ccccc}
    0 & 0 & -m(a^2\!+\!b^2) & mac & mbc \\
    0 & 0 & 0 & 0 & 0 \\
    0 & 0 & m(a^2\!+\!b^2) & -mac & -mbc \\
    0 & 0 & -I'ac & mb^2\!+\!I'c^2 & -mab \\
    0 & 0 & -I'bc & -mab & ma^2\!+\!I'c^2
  \end{array}\right),
\end{align*}
where $I' = ml^2 - I$ and $q=(\psi, \phi, \theta, x, y)$. Similarly to
the Chaplygin sleigh~\S\ref{sec:cs},
since the mass matrix is constant, the discrete dynamics is updated
explicitly through
$v_{k+\frac{1}{2}} = (\text{Id} - M^{-1}\mathcal Q(q_k)^T M)v_{k-\frac{1}{2}}$.

\begin{figure}[h]
  \begin{center}
    \includegraphics[width=2.4in]{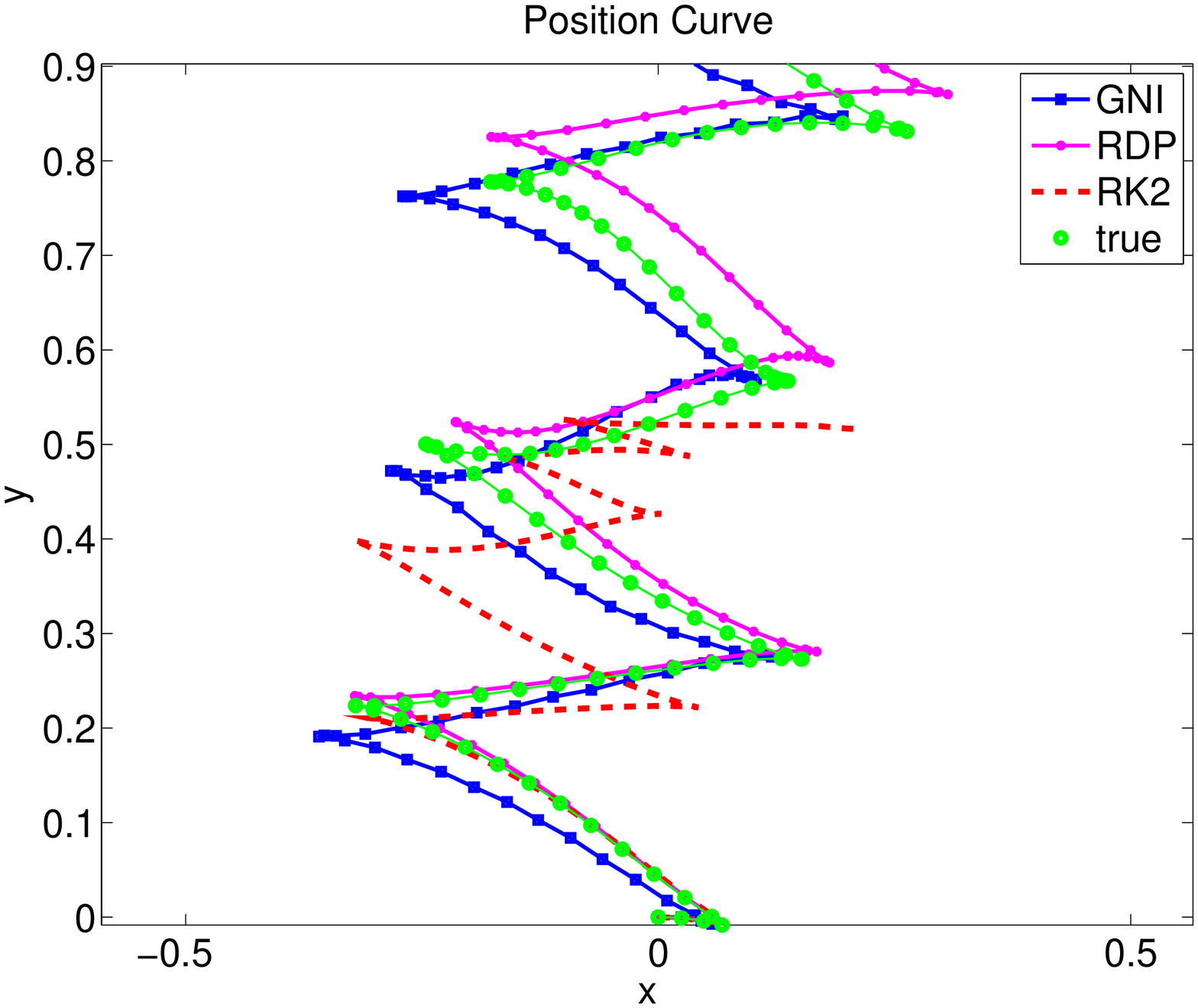}\includegraphics[width=2.4in]{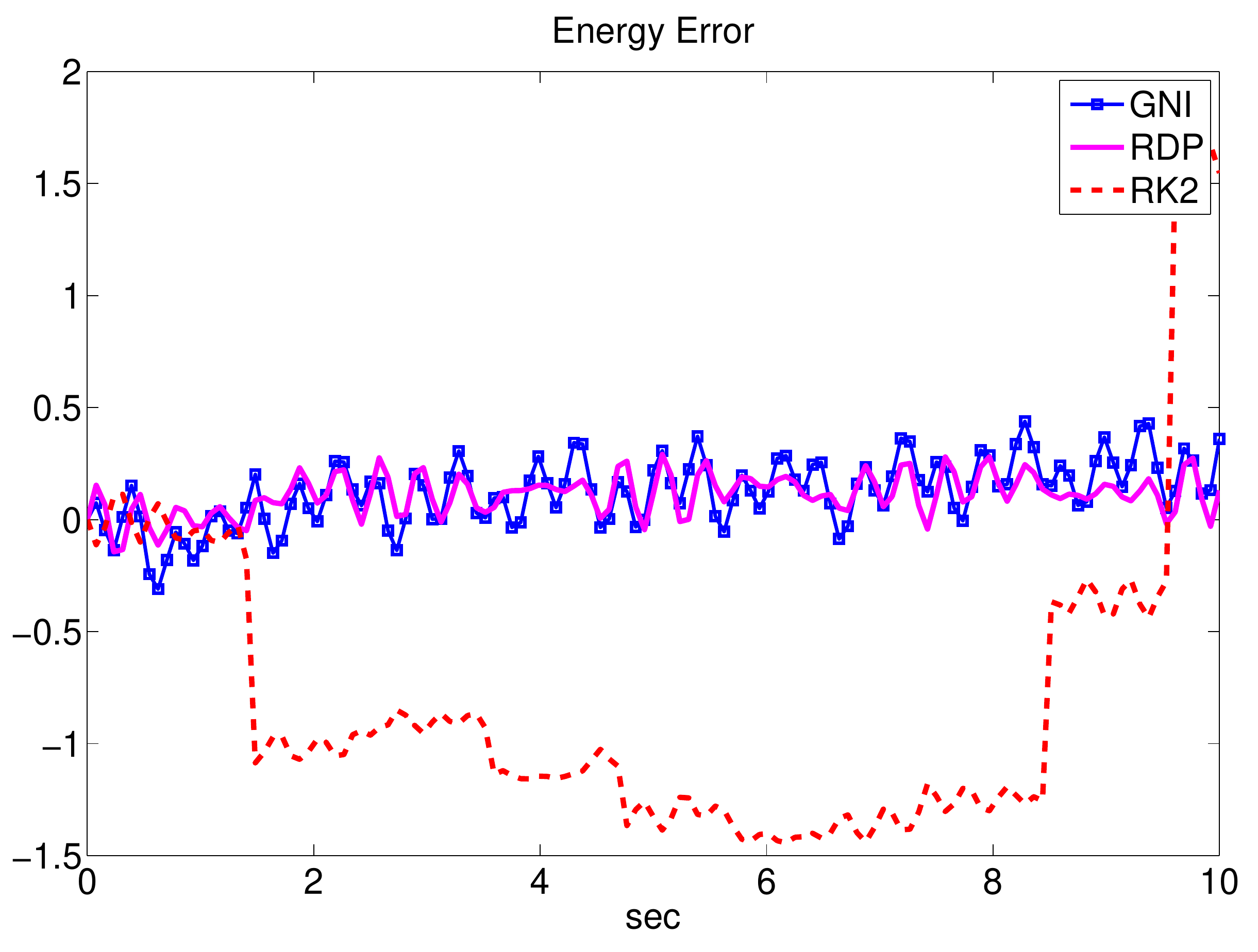}
  \end{center}
  \caption{Snakeboard integrator numerics with $N=128$ timesteps
    over the integration horizon $T=10$ sec. At such coarse resolution RK2
    method fails to follow the true trajectory while GNI and RDP have
    qualitatively correct behavior (position curve on left). One
    likely explanation lies in their better energy behavior (shown on
    right).
  }
  \label{fig:sb_plots}
\end{figure}

\paragraph{RDP Integrator.} The reduced discrete equations of motion
will be derived by substituting the Lagrangian and the connection of the
snakeboard into~\eqref{eq:rdp_dyn} and choosing the map
$\tau=\exp$. Since, particularly for the snakeboard, $\mathfrak s$ is one
dimensional and $\mathcal{A}(r)\cdot\delta$ is parallel to $e_1(r)$
for any $\delta \in T_r M$ the discrete dynamics simplifies
(see~\cite{KoMaSu,KoCrDe2009}) to
\[
\left\langle p_k - p_{k-1}, e_1(r_k)
\right\rangle = 0,\qquad \partial_u \ell_{k+\alpha} - \partial_u
\ell_{k-1+\alpha} = 0,
\]
where\vspace{-5pt}
\[
p_k= (ml^2\xi_k^1 + Iu^{\phi}_k, m\xi^2_k, 0), \qquad \partial_u
  \ell_k = (I(u^{\psi}_k + \xi^1_k), 2Ju^{\phi}_k),
\]
and the dynamics is derived by expressing $\xi_k = \Omega_k -
\mathcal{A}(r_{k+\alpha})\cdot u_k$ in terms of $r_k=(\psi_k,
\phi_k)$, $u_k=(u^{\psi}_k, u^{\phi}_k)$, and $(p_1)_k$. Note that
the discrete dynamics is linear in the unknowns $u_k$ and $(p_1)_k$
and results in an efficient explicit integrator. The reconstruction
equations are
\[
  g_{k+1} = g_k\exp ( h\xi_k), \qquad r_{k+1}-r_k = h u_k.
\]\vspace{-10pt}
\vspace{-10pt}
\paragraph{Numerical Behavior.} The studied snakeboard integrators are
second-order methods. Their advantage over similar methods is shown
through comparison to a typical second order Runge-Kutta method as
well as to the actual true trajectory. Fig.~\ref{fig:sb_plots} shows a
trajectory with initial
conditions $\psi(0)=\pi/2$, $\phi(0) = \pi/3$, $p_1(0) = -1$,
$\dot\psi(0) = 2.5$, $\dot\phi(0) = -0.02$, $\theta(0) =
0$. Sinusoidal control inputs $u_\psi = \cos(20\pi t)$, $u_\phi=\sin(2\pi
t)$ at the joints were used to create parallel parking maneuvers with
cusp points. The  CPU run-times of the compared methods are nearly
identical and are not included in the plots.

In special cases, for particular combinations of initial conditions and
inertial parameters, the GNI integrator has shown non-physical
oscillatory behavior. While this issue is most likely related to
instabilities known to occur in projection-based methods, the exact
cause remains to be determined in future work.

\section{Conclusion}

In this paper we have compared two geometric integrators for
nonholonomic dynamics, the so-called GNI and RDP integrators.
Both are constructed using differential geometric tools developed by
geometric mechanics community through a careful study of nonholonomic
dynamics during the last twenty years.
This paper shows the importance of combining different research areas
(differential geometry, numerical analysis and mechanics) to produce
methods with an extraordinary qualitative and quantitative behavior.

Such issues raise a number of future work directions. We therefore
close with some open questions:
\begin{itemize}
\item Given one of the nonholonomic integrators (GNI or RDP),  does
  there exist, in the sense of backward error analysis, a continuous
  nonholonomic system, such that the discrete evolution for the
  nonholonomic integrator is the  flow of this  nonholonomic system up
  to an appropriate order?
\item  Is it possible to use the nonholonomic Hamilton-Jacobi theory
  recently developed \cite{ILM,LeMaMa} for the construction of these
  methods or new ones?
\end{itemize}
These questions  will be part of the future work that we will develop
in the next years.

\appendix{
  \section*{Appendix}
  \section{Retraction map tangents}\label{app:dtau}
  The two common choices for retraction maps are the exponential
  map $\tau=\exp$ and the Cayley map $\tau=\operatorname{cay}$. In
  this section we provide their right-trivialized tangents
  $\operatorname{d}\tau$ of these maps and their inverses
  $\operatorname{d}\tau^{-1}$ (see~\cite{BoMa2009} for more details).

  \subsection{Exponential map}
  The right-trivialized derivative of the map
  $\exp$ and its inverse are defined as
  \begin{align}\label{eq:dexp}
    \operatorname{dexp}_x y =
    \sum_{j=0}^\infty\frac{1}{(j+1)!}\operatorname{ad}_x^jy,\quad
    \operatorname{dexp}^{-1}_x y = \sum_{j=0}^\infty\frac{B_j}{j!}\operatorname{ad}_x^jy,
  \end{align}
  where $B_j$ are the Bernoulli numbers. Typically, these expressions
  are truncated in order to achieve a desired order of accuracy. The
  first few Bernoulli numbers are $B_0=1$, $B_1=-1/2$, $B_2=1/6$,
  $B_3=0$ (see \cite{CaOw2003,Hair} for more details).
  \subsection{Cayley map}
  The derivative maps become (see~\cite{Hair} for derivation)
  \begin{align}\label{eq:dcay}
    \operatorname{dcay}_x
    y=\left(\operatorname{I}-\frac{x}{2}\right)^{-1}y\left(\operatorname{I}+\frac{x}{2}\right)^{-1},\quad\operatorname{dcay}^{-1}_x
    y=\left(\operatorname{I}-\frac{x}{2}\right)y\left(\operatorname{I}+\frac{x}{2}\right).
  \end{align}

  \section{Retraction Maps on $\mathrm{SE}(2)$}%
  \label{app:se2}%
  The coordinates of $\operatorname{SE}(2)$ are $(\theta, x, y)$ with
  matrix representation $g\in \operatorname{SE}(2)$ given by:
  \begin{align}\label{eq:se2_g}
    &g = \left[\begin{array}{ccc}
        \cos\theta & -\sin\theta & x \\
        \sin\theta & \cos\theta & y  \\
        0 & 0 & 1
      \end{array}\right].
  \end{align}
  Using the isomorphic map $\widehat \cdot: \mathbb{R}^3\rightarrow
  \mathfrak{se}(2)$ given by:
  \[ \widehat{v}= \left[\begin{array}{ccc}
      0 & -v^1 & v^2 \\
      v^1 & 0 & v^3 \\
      0 & 0 & 0
    \end{array}\right] \text{ for } v= \begin{pmatrix}
    \!v^1\! \\ \!v^2\! \\ \!v^3\! \end{pmatrix} \in \mathbb{R}^3,\]
  $\{\widehat e_1,\widehat e_2, \widehat e_3\}$ can be used as a basis
  for $\mathfrak{se}(2)$, where $\{e_1,e_2,e_3\}$ is the standard basis
  of $\mathbb{R}^3$.

  The two maps $\tau:\mathfrak{se}(2)\rightarrow \operatorname{SE}(2)$
  are given by
  \[\exp(\widehat v)\!=\!\!\left\{\!\!\begin{array}{cc}
      \!\!\left[\!\!\begin{array}{ccc}
          \cos v^1\!\!&\!\! -\sin v^1\!\!&\!\! \frac{v^2\sin v^1 - v^3(1-\cos v^1)}{v^1}\\
          \sin v^1 \!\!&\!\! \cos v^1 \!\!&\!\! \frac{v^2(1-\cos v^1) + v^3\sin v^1}{v^1}\\
          0 \!\!&\!\! 0 \!\!&\!\! 1
        \end{array}\!\!\right]\!\!&\!\!\text{if }v^1 \neq 0 \\
      \left[\begin{array}{ccc}
          1 & 0 & v^2\\
          0 & 1 & v^3\\
          0 & 0 & 1
        \end{array}\right]\!\!&\!\!\text{if }v^1=0
    \end{array}\right.\]

  \[\operatorname{cay}(\widehat v)\!=\!\!
  \left[\!\!\begin{array}{ccc} \frac{1}{4+ (v^1)^2}
      \left[\!\!\begin{array}{ccc}
          (v^1)^2\!-4 & -4 v^1 & -2v^1 v^3 + 4v^2\\
          4 v^1 & (v^1)^2\!-4 & 2v^1 v^2 + 4v^3
        \end{array}\!\!\right] \\
      \hspace{-10pt}\begin{array}{ccc} 0 \qquad &\qquad 0 \qquad &\qquad 1 \end{array}
    \end{array}\!\!\right]
  \]

  The maps $[\operatorname{d\tau}_{\xi}^{-1}]$ can be expressed as the
  $3\times 3$ matrices:
  \begin{align} [\operatorname{dexp}^{-1}_{\widehat v}] \approx
    \mathbf{I}_3 - \frac{1}{2} [\operatorname{ad}_v] +
    \frac{1}{12}[\operatorname{ad}_v]^2, \label{eq:se2_dexpinv}\\
    [\operatorname{dcay}^{-1}_{\widehat v}] = \mathbf{I}_3
    -\frac{1}{2}[\operatorname{ad}_v] +
    \frac{1}{4}\left[\begin{array}{cc} v^1 \cdot v & \mathbf{0}_{3\times
          2}\end{array}\right] \label{eq:se2_dcayinv},
  \end{align}
  where
  \[ [\operatorname{ad}_{v}] = \left[\begin{array}{ccc}
      0 & 0 & 0 \\
      v^3 & 0 & -v^1 \\
      -v^2 & v^1 & 0
    \end{array}\right]. \]
}

\end{document}